\documentclass[runningheads]{llncs}
\usepackage[hyphens]{url}
\usepackage{graphicx}
\usepackage{mdframed}
\usepackage{framed}
\usepackage{setspace}
\usepackage{parcolumns}
\usepackage{listings}
\usepackage{comment}
\usepackage{microtype}
\usepackage{caption}
\usepackage{graphicx}
\usepackage[hidelinks,breaklinks=true]{hyperref}
\usepackage[hyphenbreaks]{breakurl}
\newcommand{\secref}[1]{\S \ref{#1}}

\begin{document}
\title{One Protocol to Rule Them All? \\On Securing Interoperable Messaging}

\author{
Jenny Blessing\inst{1} \and
Ross Anderson\inst{1,2}}

\institute{University of Cambridge \and University of Edinburgh\\
\vspace{.7em}
\email{\{first.last\}@cl.cam.ac.uk}}

%
%
%
%
%
\maketitle              

\begin{abstract}

European lawmakers have ruled that users on different platforms should be able to exchange messages with each other. Yet messaging interoperability opens up a Pandora's box of security and privacy challenges. While championed not just as an anti-trust measure but as a means of providing a better experience for the end user, interoperability runs the risk of making the user experience worse if poorly executed. There are two fundamental questions: how to enable the actual message exchange, and how to handle the numerous residual challenges arising from encrypted messages passing from one service provider to another -- including but certainly not limited to content moderation, user authentication, key management, and metadata sharing between providers. In this work, we identify specific open questions and challenges around interoperable communication in end-to-end encrypted messaging, and present high-level suggestions for tackling these challenges.



\end{abstract}

\section{Introduction}
Users of end-to-end encrypted (E2EE) messaging services have long existed in a world where they need to use the same service as another user in order to communicate. A Signal user can only talk to other Signal users, an iMessage user can only use iMessage to communicate with other iPhone users, and so on. Platform interoperability promises to change this: the vision is that a user of a messaging service would be able to use their platform of choice to send a message to a user on a different service — following the precedent of email and SMS. Proponents of this kind of open communication have argued that it will benefit both the end user and the market for services. If users can message each other using their preferred service, they can enjoy their user experience of choice, and if there is less pressure to use a service simply because others use it, this can eliminate network effects and market monopolies.

An interoperability mandate for end-to-end encrypted messaging systems is no longer hypothetical: the European Union’s Digital Markets Act (DMA) came into force in November 2022, and Article 7 requires that the largest messaging platforms (termed “gatekeepers” by the DMA) allow users on smaller messaging platforms to communicate directly with users on the large platforms~\cite{dma}.\footnote{The European Commission will not officially designate which platforms will be considered gatekeepers (and therefore fall under the mandate) until September 2023.} The mandate applies only to the gatekeepers, with any non-gatekeeper platforms free to choose whether they wish to interoperate with other platforms. In accordance with the DMA, the gatekeepers cannot deny any ``reasonable'' request. Notably, it leaves the technical implementation details for the platforms to determine. In the U.S., the 2021 ACCESS Act proposed similar requirements but has yet to make any headway in Congress.

In this paper, we survey and explicitly articulate the security and privacy trade-offs inherent to any meaningful notion of interoperability, focusing primarily on the supporting aspects of E2EE communication which are largely agnostic to the actual method of message exchange. Designing a system capable of securely encrypting and decrypting messages and associated data across different service providers raises many thorny questions and practical implementation compromises.\footnote{Messaging interoperability raises numerous systems challenges, such as the latency impact of transferring messages between service providers when users expect communication to be instantaneous, particularly with audio/video calls, impact of bridging on mobile device battery life, etc. We limit our scope to challenges with direct security impact.}

We outline what current solutions exist and where existing protocols fall short, and propose high-level solutions for tackling some of these challenges. For the sake of the discussions that follow, we assume platforms make a genuine effort to develop a system that emphasizes security and usability, though in practice they may degrade the user experience for interoperability, whether as a matter of necessity (WhatsApp has more features than Signal) or choice (to maintain some degree of customer lock-in). But as we will discuss, these challenges exist even if platforms make real efforts to open up their systems.

We argue that while it is possible to achieve interoperable end-to-end encrypted communications with contemporary messaging services, this will require numerous new protocols and processes, both cryptographic and human, to maintain reasonable levels of security and usability. The conceptual simplicity of messages passing back and forth between services belies the difficulty of the problem. Interoperability doesn't just mean co-opting existing cryptographic protocols so that one service provider can pass messages along to another -- it encompasses the many supporting features and protocols that make up contemporary E2EE applications. The resulting complexity of the system may inherently compromise the level of security due to the increased number of moving parts, just as key escrow mechanisms endanger cryptography even if the escrow keys are kept perfectly secure. 

The DMA includes a purported safeguard that the ``level of security, including the end-to-end encryption, must be maintained" in an interoperable service, but this raises as many questions as it answers. ``Level of security" goes well beyond the mere fact of using an end-to-end key exchange protocol. A platform may use a proprietary E2EE protocol that does not provide forward secrecy, a de facto standard in preventing compromise of past communications~\cite{borisov2004off}, does not regularly rotate encryption keys, or neglects various other cryptographic guarantees. Until recently, the Swiss messaging app Threema openly had no forward secrecy at the E2EE layer \cite{threema_ibex}, a design which led to multiple security problems~\cite{patersonthree}. Will these be considered valid reasons for a gatekeeper to deny a request to interoperate?

How will a gatekeeper verify the requesting service’s encryption protocol, along with the authentication and content moderation schemes used? Will they need to take the requesting service’s word for it, or else invest the resources to do a proper security audit? If the requesting service is closed-source, will the gatekeeper have a right to request access to the other's source code should they wish to do an audit? All widely-used E2EE messaging services have significant differences in both protocol and implementation, including fundamentally different design decisions impacting security and usability \cite{ermoshina2016end}. For instance, WhatsApp and Signal, despite both being based on the Signal protocol, handle key changes when a message is in flight differently~\cite{isoc2022dma}. When a recipient's keys change in the course of a conversation (e.g., because they uninstalled the app), Signal discards any messages sent after the change, while WhatsApp chooses to deliver them once the recipient comes back online. Both designs are legitimate~\cite{tufekci2017response,moxie2017whatsapp}.

The greatest challenges are non-technical: interoperability will require competing platforms to cooperate and communicate, both in the current design phase and after any agreed-upon interoperable scheme has been deployed. Each service provider will need to trust the others to provide authentic key material, enforce certain spam and abuse policies, and generally to develop secure software with minimal bugs. A vulnerability or outage in one service now propagates to all other services with which it interoperates. Existing cryptographic protocols mitigate, but do not eliminate, these trust requirements. There is simply no getting around the fact that interoperability represents a dramatic expansion in the degree of trust a user will need to place not only in their own messaging service but also in any used by their communication partners, a point to which we will return throughout the paper. Much of the rhetoric around messaging interoperability at a recent European Commission-hosted stakeholder workshop~\cite{ec2023stakeholders} and elsewhere calls to mind the quip ``if you think cryptography is your solution, you don't understand your problem.''\footnote{This quotation has been alternately attributed to Roger Needham, Butler Lampson, and Peter Neumann~\cite{anderson2020security,kolata2001neumann}, and paraphrased by Phillip Rogaway~\cite{rogaway2015moral}.}

As service providers begin to make plans to comply with the DMA by the March 2024 deadline, it is important to tackle open questions now that providers are grappling with them, rather than waiting until platforms have invested a major effort in developing new systems and protocols. The stated policy goals of interoperability are to improve the user experience and decrease network effects, yet a poor execution risks doing the opposite. Open communication by itself achieves little if it undermines the very reasons people use end-to-end encrypted messaging in the first place.


\section{Implementation Paths}
There are two broad strategies to enable separate messaging platforms to talk to each other~\cite{berec2022report,hodgson2023blog}: either all platforms adopt a common communications protocol (native interoperability), or each platform publishes an open API allowing others to communicate with them through a bridge. There are more flavors than these, including many hybrid possibilities with varying levels of implementation and maintenance feasibility---a platform could, for instance, support multiple protocols.

\subsection{Standard Protocol}
There are several existing candidates for selecting a universally adopted end-to-end key establishment protocol. The Matrix Foundation has developed the federated and interoperable Matrix protocol \cite{matrix_spec}. The Signal protocol (formerly TextSecure) has been around for roughly a decade now and is the only open-source E2EE protocol that has been deployed at a scale of billions of users. More recently, the IETF standardized a new E2EE message exchange protocol, MLS (Messaging Layer Security) \cite{mls_ietf}, which is intended to provide more efficient group communications than in Signal and related protocols. In February 2023, the IETF created a new ``More Instant Messaging Interoperability'' (MIMI) working group, the IETF's latest messaging interoperability standardization effort, dedicated to establishing the ``minimal set of mechanisms'' needed to allow contemporary messaging services to interoperate~\cite{ietf_mimi}. Among other aspects of messaging standardization, MIMI will seek to to extend MLS to deal with user discovery (``the introduction problem'') as well as content formats for data exchange~\cite{ietf_mimi}. Work is ongoing and expected to continue well into 2024.

Any choice here is not obvious.\footnote{See \url{https://xkcd.com/927/.}} Several of the largest messaging services already use variations of the Signal protocol, and Meta explicitly advocated for widespread adoption of Signal at the European Commission's stakeholder workshop~\cite{ec2023stakeholders}. Another possibility is some sort of hybrid option: Matrix has proposed Matrix-over-MLS as part of the IETF's MIMI working group \cite{ralston2022ietf}. But any agreed-upon standard would eventually have to support many of the features and functionalities of all widely used E2EE applications.

Switching to a standard communications protocol poses immense challenges given the variety of protocols currently in use. Signal, WhatsApp, Viber, Facebook Messenger, and others rely on some variation of the Signal protocol, though they have developed different implementations (and, particularly in the case of group communications, different protocol versions). Telegram, Threema, and iMessage use custom protocols for the end-to-end encrypted layer, with Threema even using a custom client-to-server protocol~\cite{unger2015sok,threema_whitepaper}. Element uses the Matrix protocol~\cite{matrix_spec}. Existing messaging services would need to either switch to a common protocol or support multiple protocols. Service providers not only have to agree on a single protocol, but in many cases would have to redesign their entire system and manage a major migration to the new interoperable version.

There are also valid concerns around hampering innovation: Moxie Marlinspike, one of the co-creators of the Signal protocol, has argued that centralized, unfederated protocols evolve more rapidly than decentralized ones, where the latter have to deal with a wide diversity of clients, implementations, and deployments~\cite{marlinspike2019ecosystem_talk,marlinspike2016ecosystem_blog}. For these reasons and others, a common protocol seems less practical in the near future than client-side APIs, not least because of the time constraints imposed by the DMA.

\subsection{Client-side Bridges}
In the face of substantial political, economic and technical obstacles to a universal communication standard, the developer community has begun to gravitate towards the idea of providing interoperability via public APIs, at least in the short term~\cite{matrix_blog}. Each service provider could largely keep their existing E2EE protocol and implementation, but would provide a client-side interface to allow other messaging services to interact via a bridge between the two services.

Depending on the precise architectural design, such an interface would entail decrypting messages locally on the recipient’s device after they are received from the sender's service provider, and then re-encrypting them with the recipient service provider's protocol. If the bridge (and therefore the key establishment protocol) runs on the client, this does not break the general notion of end-to-end encryption, at least theoretically. Only the endpoints (the client devices) see the plaintext message; no platform server is able to decrypt messages at any point. While running a server-side bridge is a technical possibility, both Meta and Matrix have acknowledged that server-side bridging has been ``largely dismissed'' since it would require message decryption and re-encryption in view of the service (violating the core principle of end-to-end encryption)~\cite{hodgson2023dma_workshop}. A hybrid of the two, in which only the E2EE protocol is run in a client-side bridge with a server-side bridge doing much of the actual message transport, is also a possibility to handle systems challenges that may arise from asking the client to do too much work.

Some service providers already offer similar interfaces, either to integrate with other services or to allow third-party clients access. iMessage, Apple's end-to-end encrypted messaging service which interoperates with SMS, provides an example of client-side bridging at scale. It is well worth noting, however, that the technical feasibility of the message exchange, even at large scale, does not mean that this can be smoothly applied to interoperating E2EE services. SMS suffers from serious privacy issues, as all messages are unencrypted, and interface design choices (namely, bubble color) have led to widespread aversion to SMS conversations. We will return to both points in greater depth later on.

\hfill\break
Both of these approaches are far from a panacea. Eric Rescorla has written and spoken at length about the challenges of writing, standardizing, and implementing protocol specifications~\cite{rescorla2023architecture}. Client-side bridging comes with its own set of challenges: each service requesting access will need to build a different bridge for each provider with whom they want to interoperate. This has inherent security implications arising from the amount of excess code a service would need to add to achieve interoperability with a meaningful number of services. And while client-side bridging may not break E2EE in the technical sense that the plaintext messages are only viewable on the client's device, it is indisputably not a conventional meaning of E2EE. Moreover, providing \textit{an} interface is not the same as providing a \textit{usable} interface, which is a challenge even when a provider is giving their best effort. To aid in providing a reasonable developer experience, Rescorla has suggested a set of main interface requirements, including unambiguous interface specifications, stable interfaces, test servers, and real-time support from engineers at the other service~\cite{rescorla2023architecture}. Needless to say, these are far easier said than done.

\vspace{.7em}
\noindent\textbf{Restrictions on API access:} In practice, a service provider's interface cannot truly be ``open" due to the sensitivity of the data being exchanged. Large service providers will individually approve requests by smaller service providers. The process for filtering requests may vary by provider, though they cannot deny a ``reasonable" request. An interoperable interface will likely use some sort of revocable access key to manage access. Providers then need to figure out how to manage key requests and key storage, including some sort of service-level identity indicator \cite{isoc2020whitepaper}.

But API request filtering is not merely a gatekeeping measure: platforms need to be able to detect and block bulk spam and forwarding services in real time. WhatsApp relies heavily on behavioral features to detect spam clients at the time of account registration in its existing public-facing interfaces, stating that the majority of use cases of these interfaces are spam \cite{jones2017whatsapp}. Will Cathcart, the current head of WhatsApp, identified the need to restrict or outright block certain services as one of the most important considerations under an interoperability mandate \cite{newton2022cathcart}.

The EFF has raised the spectre of a malicious actor creating a fake messaging service with a number of fabricated users and requesting access \cite{cyphers2021eff}. Large service providers will need to set certain objective and justifiable thresholds that give them the flexibility to respond only to legitimate requests. This is seemingly within the bounds of the DMA, though of course this will depend on what types of requesting services the European Commission considers to endanger the ``integrity'' of a service~\cite{dma}.

Many of the data sharing and privacy concerns surrounding third-party access are not fundamentally new: Facebook’s infamous Cambridge Analytica scandal was made possible through Facebook’s external app API. What has changed with the DMA is that providers have far less flexibility to refuse or revoke access, let alone in real time as a situation is unfolding. Service providers will need a fair amount of latitude in their ability to deny access requests to continue to guard against malicious data scraping and mining, regardless of whether interoperable message is implemented through client-side bridging or an open standard.

\section{Open Challenges}

The security and privacy considerations associated with trying to reconfigure end-to-end encrypted systems to communicate with each other are too numerous for us to attempt complete coverage. We focus on five general areas in particular: user identity, key distribution, user discovery, spam and abuse, and interface design. We try to keep the discussion high-level so that it is largely agnostic to the message exchange architecture (i.e., whether messaging platforms have adopted a standard protocol or opted for client-side bridges).

\subsection{User Identity}
End-to-end encryption is meaningless without sufficient verification of the authenticity of the ends. There are two layers to identifying users:
\begin{enumerate}
    \item \textit{Cryptographic Identity:} First, how do you know whether a given public identity key belongs to a user Alice's account?
    \item \textit{Real-world Identity:} Second, once you have verified the public key attached to Alice's account on this service, how do you know that the user ``Alice Appleton" who has contacted you is the Alice Appleton you know in real life?
\end{enumerate}

Both cryptographic identity and real-world identity are needed to assure a user that they are indeed talking to the right person. We discuss each of them in turn.

\subsubsection{Cryptographic Identity:}
A user's public key forms their ``cryptographic identity''. Currently, each messaging provider maintains their own separate public key directory for their userbase. When Alice wants to talk to Bob on a given messaging application (which both Alice and Bob use), Alice's client queries their provider for Bob's public key. The provider looks up Bob in their centralized key database and uses this information to establish an end-to-end encrypted communication channel between Alice and Bob.

Existing key distribution protocols generally require users to trust the provider to store and distribute the correct keys, and are vulnerable to a malicious provider or compromised key server. Interoperability further complicates trust establishment, since users will now have to place some degree of trust in a separate service provider. We revisit this question in greater depth in \secref{section:key_distribution}.

\subsubsection{Real-world Identity:}
Tying a cryptographic identity to a real-world identity is an even more challenging problem since security and privacy are somewhat in conflict here. Messaging services vary in the information they ask of users: WhatsApp and Signal both require users to provide a phone number at the time of account registration, though Signal is currently working on username-based discovery so that a user's phone number is known only to Signal \cite{whittaker2022tweet}. iMessage uses the email address associated with a user's Apple ID by default, but can also be configured to identify a user by their phone number. The Swiss messaging app Threema, on the other hand, identifies users through a randomly generated 8-digit Threema ID, and does not require a phone number or any other information tying a user account to an real-world identity scheme \cite{threema2019anonymity}. How is a WhatsApp user to know the Threema user is who they claim to be? At the moment, the only option would be to verify identities through an out-of-band channel (e.g., SMS, email, or meeting in person), which users rarely do in practice~\cite{vaziripour2017you,vaziripour2018action,schroder2016signal}. And even if an identity assurance ceremony is performed once in person, many vectors of account takeover have been industrialised by the cybercrime community (such as SIM swapping) while others are widely available to state actors (such as SS7 hacking).

While a real-world identity mechanism like a phone number or email address may give some identity assurance, there are many valid reasons why a user might not want to tie their identity on a messaging service to their real name, so the use of handles or pseudonyms is a desirable property for some messaging platforms to offer. Even in the absence of privacy concerns, users need different handles to cope with congested global namespaces; the `ross.anderson' at the Cambridge Computer Lab becomes `rossjanderson' on Twitter and `profrossanderson' on Threads.

Given the importance of identity assurance to interoperable communication, however, it may be that identity assurance and anonymity cannot be reconciled absent out-of-band user verification. Under the DMA, would a service provider be allowed to reject a request for interoperability from a messaging service that does not collect some form of external identity from its users? It is difficult to argue that a service's ``level of security'' is maintained if platforms are obligated to interoperate with a service that does not rely on some suitably accredited external identity scheme. And if so, would this give smaller services an incentive to remove support for pseudonymous account registration? Such a decision may interact with other EU provisions around identity (such as eIDAS, the European Union's electronic identity verification service) as well as broader policy questions (such as identity escrow and age verification). The eventual outcome might be a creeping `real names' policy, to the disadvantage of users with a genuine reason to seek anonymity -- from political dissidents and stigmatised minorities to survivors of intimate abuse. Such users might have to seek out grey market service providers that have opted not to interoperate with regulated platforms, which would be marginalized and perhaps pass some stigma to their users.

Importantly, one of the main attractions of these platforms is that it is simple to sign up for them. In most cases, user identity is bootstrapped off of a user's phone number, where a provider sends the user an SMS message with a random string which the user then enters in the app to prove control of said phone number. This ease of use must be preserved in any interoperable scheme, for instance by using an identifier the user already has memorized.


\subsection{Key Distribution}
\label{section:key_distribution}

Contemporary end-to-end encrypted messaging systems maintain platform-specific, centralized key directories to store their users' encryption keys. When a new user registers an account, the user's device generates several public-private key pairs (the precise number and type of key pairs depend on the protocol used) and sends the public keys to the service provider to store in its directory.

From a security standpoint, this reliance on the service provider to store and distribute encryption keys is a major weakness in existing systems. Since the service provider controls the public key directory, a malicious provider could compromise end-to-end security by swapping a user's public key for one under their control, either of their own volition or because they were legally compelled or otherwise pressured to do so. The confidentiality of the communications, then, hinges on trusting that the service provider has provided the correct keys. Existing protocols fail to prioritize this: MLS does not tackle the storage or distribution of keys, only how they might be used to send and receive cross-platform messages.

Interoperable communication further complicates trust issues around key storage and distribution. How would one service provider share the identity keys of its users with another in a way that satisfies user privacy expectations? And how can one platform be certain that the other has shared the correct keys? A platform could conceivably trick a user of another service into talking to a different person than the one with whom they believe they are communicating. Each service provider has no choice but to trust the others. Any open communication ecosystem will need to design explicit and effective controls around accessing, sharing, and replacing keys. Users' identity keys change constantly—every time they delete and reinstall the app or get a new phone, their device generates a new set of keys and the provider updates their directory accordingly. Service providers will need an efficient way of conveying user key changes to other providers. There is some ongoing academic work to develop a decentralized key infrastructure based on existing digital ID systems~\cite{turing_trustchain}, but these are still in the proof-of-concept stage.

\subsubsection{Key Transparency:}

Key transparency is an area of active research that, in theory, would eliminate the need to trust the service provider to share the correct keys. The general idea is that a service provider will still maintain a large key directory, but this directory is now publicly accessible and auditable by independent parties (either a third-party or possibly the service providers themselves auditing each other). A provider cryptographically commits to a user’s identity to publicly key mapping at the time of key generation (such that it cannot be changed without detection), and further periodically commits to the full directory version. Users (more precisely, users' messaging clients) can then verify both their own keys as well as their contacts' keys to detect a provider serving different versions of its key directory to different users. In practice, service providers would likely maintain separate key directories, but other providers would be able to query this directory in a privacy-preserving manner such that an external provider could only query for individual users, along with other privacy mitigations. Note that an auditable keystore system does not \textit{prevent} a key-swapping attack from taking place, it only \textit{detects} when such an attack has happened after the fact.

CONIKS~\cite{melara2015coniks}, the first end-user key verification design, was proposed in 2016 but suffered from scalability issues due to the frequency with which users would need to check that their key is correct. Several other key transparency systems have been proposed and, in some cases, deployed, in the years since. Subsequent academic research has built on CONIKS, formalizing the notion of a verifiable key directory and mitigating scalability concerns by making the frequency of user key checks depend on the number of times a user's key has changed, along with providing a handful of other practical deployment improvements \cite{chase2019seemless,malvai2023parakeet}. For various practical reasons, the largest industry providers have been slow to deploy key transparency. Google released a variation of CONIKS that enables users to audit their own keys in 2017~\cite{google2017transparency}. More recently, in April 2023 Meta announced plans to roll out large-scale key transparency across WhatsApp~\cite{whatsapp_transparency}, though there are still many unresolved implementation questions such as how public audits will be carried out and who the auditors will be.

But deploying a verifiable key directory service at scale across multiple service providers large and small is another matter. The authors of CONIKS neatly outlined several of the main barriers to deployment back in 2016 based on discussions with engineers at Google, Yahoo, Apple, and Signal \cite{melara2016blog}. The most serious challenges boil down to the difficulty of distinguishing between legitimate and adversarial key changes. When Alice gets a new phone or forgets her iMessage password and resets her keys, how can Alice's service provider convince other providers accessing and auditing the key directory that they have legitimately identified Alice through some real-world identity mechanism, and that these new keys do in fact still belong to Alice? Will Alice need to somehow prove her identity not just to her own service provider, but to all other providers? We are not aware of any existing key transparency proposals that solve these problems adequately. In the absence of genuine end-to-end identity verification, will WhatsApp simply have to take Telegram's word for it?

And completely apart from the technical challenges, we cannot assume that platforms will agree on how such a key management service should work. Messaging providers have already made different design decisions around detecting false positives and when to send a user a security warning that one of their contacts' keys has changed, usually in an attempt to avoid inundating users with security warnings. Until very recently, iMessage did not even provide users with an option to manually verify their keys. Their new Contact Key Verification system claims to alert users if an ``exceptionally advanced adversary, such as a state-sponsored attacker'' has managed to secretly join a chat~\cite{apple2022security}, but few technical specifics of how this works on the backend are publicly available. Many other widely used E2EE services inundate the user with in-chat notifications every time one of their contacts' keys has changed. The complexity of a distributed key verification service, however, makes it even more likely to issue false security warnings due to various synchronization problems.


\subsection{User Discovery}
We can now build on the above discussions of user identity and key management to consider how the process of learning which service(s) a user uses and/or prefers might work.

There are two separate but related design principles, advocated by multiple NGOs, that should be considered in the development of any discovery mechanism~\cite{cdt2022journeys,isoc2022dma}:
\begin{enumerate}
    \item \textbf{Separate Communications:} Users must be able to keep their communications on different messaging apps separate if they choose. The Center for Democracy \& Technology offered the analogy of using a work email and a personal email, a paradigm adopted by the vast majority of the general public \cite{cdt2022journeys}. Prior work has shown that some users likewise use different messaging apps for different purposes~\cite{nouwens2017whatsapp,arnold2020interoperability,griggio2022caught,wik2022interoperability}, a feature that should be preserved in an interoperable world. In other words, users should retain the ability to sign up for a messaging service and opt to receive messages from users on that same service only.
    \item \textbf{Opt-Out by Default:} Related to the first principle, users should be opted-out of discovery by default to maintain reasonable user privacy expectations.\footnote{The discussion around opt-in versus opt-out discovery in messaging interoperability is often compared to email since email addresses are openly discoverable and contactable by anyone. But there are well-recognized social conventions and distinctions among email services that do not exist in the messaging ecosystem. For instance, given two of Alice’s email addresses, “alice.appleton@gmail.com” and “alice.appleton@company.com”, one can reasonably infer which types of communications Alice would like sent to each address without needing to discuss with Alice.} When a user signs up for a messaging service, they consent to discovery within that service---and only that service. We see two high-level designs where this could be accomplished: using service-specific identifiers (as with email), or allowing users to change their discoverability preferences in a per-service basis in the settings of the app(s) they use (i.e., Alice can choose to be discoverable on iMessage but not Telegram, etc.).
\end{enumerate}

Keeping these principles in mind, we have two general models for forming user identifiers \cite{rosenberg2022taxonomy}:
\begin{enumerate}
    \item \textbf{Service-Independent:} Users use the same real-world identifier (e.g., phone number) across multiple services. Alice wants to contact Bob for the first time, but since Bob's identifier is not linked to any one specific service, Alice still needs to figure out where to send her message. Presumably, her messaging app will present some sort of app selection interface for Alice to choose which service to use to contact Bob.\footnote{Many interface design questions arise here. For now, we focus primarily on privacy concerns in user discovery, and revisit the user experience in \secref{section:user_interface}.} If Bob is discoverable on multiple services, either Alice asks Bob which service he prefers out-of-band (e.g., in-person, over email, etc.), or Alice simply selects one of the options from the interface based on her own preferences. If users do not want to select a service manually each time they begin a conversation, then there are various options for automation. The simplest might be for each user to have a priority list (e.g. try Signal, then iMessage, then WhatsApp), just like the negotiation of TLS ciphersuites or EMV credit card chip authentication methods. But if a user wanted family messages on WhatsApp and work messages on Signal, then this would become more complex still.

    \vspace{.5em}
    \item \textbf{Service-Specific:} Alternatively, Bob's identifier could be linked in some way to a specific service such that Alice's service provider knows where to deliver messages addressed to Bob. This is analogous to email, where each identifier is scoped to a particular namespace (i.e., alice.appleton@gmail.com and alice.appleton@cam.ac.uk do not necessarily refer to the same person). In the case of messaging, Rescorla has suggested that this could look something like ``1.415.555.0123@whatsapp.com''~\cite{rescorla2022discovery}.

    \vspace{.5em}
    Of course, asking the user to enter this mapping would be messy from a usability standpoint, and in many cases the user would still need to go through an out-of-band process to obtain the scoped identifier from the other user. To keep the user experience seamless, this identifier-to-service mapping could be hidden from the user and only used internally. 

\end{enumerate}

\subsubsection{Centralized Directory Service:} The tricky part is figuring out how each service provider learns which services are associated with a given phone number. Rescorla and others have floated the idea of a large-scale, centralized database of phone number-to-messenger mappings, similar to how the PSTN (Public Switched Telephone Network) maintains a large database mapping phone numbers to carriers~\cite{rescorla2022discovery}. At a high level, a user record would be added to the directory service or updated at the moment of app installation, once that user has gone through some real-world verification process to their service's satisfaction (e.g., provided a random code sent to their phone number). In theory, this directory might be part of a centralized key distribution service.

But creating such a database for E2EE messaging services is more complicated than the PSTN database for a number of reasons (as Rescorla acknowledges). First and foremost, there is no privacy to speak of in SMS, whether we're talking about discovery or message contents. Any carrier can query the database---which is, of course, one of the reasons SMS is overrun with spam. Second, number-to-carrier is a one-to-one mapping, while number-to-messaging service is a one-to-many mapping (assuming that a significant number of users continue to use more than one messaging service). As discussed above, there are several different design options for selecting a service, including letting the message sender select through an interface, letting the message recipient indicate a preferred service, or perhaps even asking users to provide a ranking of services by context and then choosing the highest-ranked service common to both. Each of these would require a different set of cryptographic protocols to maintain certain privacy-preserving attributes.

In contrast to the PSTN database, in which phone numbers are rarely changed or removed, users may frequently adjust their discoverability preferences for different services, for instance as they make a new acquaintance who wants to use a particular service or receive too many unwanted spam messages from a different service. We will need effective identity revocation mechanisms: if Bob changes his mind and no longer wants to be discoverable by other platforms, how can Bob’s service provider communicate this to the other platforms and gain reasonable assurance that they have actually removed Bob and all associated data from their servers?

In particular, the centralized nature of such a service raises several security and privacy concerns, some, but not all, of which can be solved using known cryptographic schemes. The ability of an individual user to query this directory and learn which apps another user is associated with is probably the least concerning since this is the case for several of the largest platforms today. On Signal, for instance, you need only know someone's phone number to learn whether they are also on Signal. While a user can now rapidly retrieve a list of apps used by someone else, instead of having to manually install and test each one, this has minimal impact on the attack surface compared to other challenges. Presumably the directory would be rate-limited to some extent to prevent mass scraping, though the success of such a mitigation will come down to how rigorously clients are identified. 

The most serious concern is that the identity service would know precisely who is talking to whom based on user identity lookups. This might be mitigated through private information retrieval (PIR) or private set intersection (PSI) techniques for anonymous contact discovery, but while academic work has made great strides in improving the scalability of PIR, it is unclear if such schemes are workable on the scale of billions of users. Any contact discovery database will need to update each time a new user joins a platform to maintain basic functional requirements of a contemporary messaging service. When Bob signs up for a new platform, he expects to be able to send and receive messages instantly, not to have to wait 24 hours until the database has been updated to include him. These requirements make PIR impractical given that current academic research relies heavily on server preprocessing of a periodically updated database.
 
A centralized identity-mapping service would also know all messaging platforms used by every individual, a fact with significant implications for user privacy. A related point is the need to make any such lookup service compatible with users' ability to opt in to interoperability on a service-by-service basis. In other words, Alice, who uses $S_A$, could opt in to discovery by users on $S_B$, but not $S_C$. Would $S_C$ still be able to access Alice's records? How could access be controlled such that each provider is only able to query a subset of user records, and who would enforce this? Perhaps instead of a universal lookup service, each set of providers that interoperates shares a lookup server with records of jointly discoverable users, though this still poses many of the same questions around trust, shared hosting responsibility, and scalability.

\subsubsection{SPIN:} To avoid having to use a centralized service, the IETF MIMI working group has introduced a high-level framework for a new identity mapping protocol called SPIN (Simple Protocol for Inviting Numbers)~\cite{mimi2022spin} that operates similarly to existing account setups. SPIN assumes that all users are identifiable by their phone number, communicating users already know each other's numbers (as WhatsApp and others operate today), and that users' devices are online at the moment the conversation is started. When Alice wants to send a message to Bob, who uses a different service, Alice's client sends an SMS message to Bob's device requesting the services Bob supports, and Bob's device replies. While this method of identity mapping avoids the problem of using a large-scale directory service, it brings its own downsides. It requires modifications to and the cooperation of the underlying mobile OS (namely, iOS or Android) in addition to the messaging platforms, requires users to be online to be discoverable, and effectively adds an extra round trip (and possibly a small delay) to the normal communication process~\cite{rescorla2022discovery}.

\subsection{Spam and Abuse}

Content moderation is a real challenge in deploying an interoperable network of networks. Detecting and handling spam and abuse effectively is an unsolved problem with plaintext content, let alone encrypted content, let alone across multiple complex distributed systems~\cite{mayer2019content,scheffler2023sok,anderson2022chat}.

\subsubsection{Existing Techniques}
\hfill\\
\vspace{-.5em}

\noindent The existing providers have spent years building up their content moderation systems, training complex machine-learning models and hiring human moderators to resolve hard cases and feed back ground truth. The scale is astonishing: WhatsApp bans nearly 100 million accounts annually for violating its terms of service~\cite{ec2023stakeholders}. It is unreasonable to expect that providers will all adopt some new universal content moderation system, unless perhaps mandated to do so by governments (more on this in~\secref{section:content_detection}).

For now, let us assume an end-to-end encrypted system where only the users can access message contents. In such a setup, providers rely on two primary techniques for content moderation: user reporting and metadata~\cite{kamara2022outside,pfefferkorn2022content}.

\vspace{.7em}
\noindent\textbf{User Reporting:} The most effective content moderation scheme at scale is user reporting~\cite{pfefferkorn2022content}. How might reports work for users communicating through two different service providers? Suppose Alice is using $S_{A}$, to exchange messages with Bob, who uses $S_{B}$. Bob sends Alice an abusive message, prompting Alice to block Bob. Either $S_{A}$ needs a way of passing this on to $S_{B}$, or $S_{A}$ continues to receive further messages from Bob but simply opts not to display them to Alice. For instance, email handles blocked users by automatically redirecting any future emails from them to a user's spam folder.

Suppose Alice also reports Bob for good measure. Presumably $S_{B}$ would be responsible for handling the report as Bob is their user, but since Alice has reported him on $S_{A}$'s interface, $S_{A}$ will need to pass along relevant information to enable $S_{B}$ to take appropriate action. When a message is reported, the clients of several E2EE services, including WhatsApp and Facebook Messenger, automatically send the plaintext of the previous five messages in a conversation to the provider along with the reported message to provide additional context~\cite{kamara2022outside}. Will this policy be maintained in an interoperable context, such that in order to report Bob, Alice ends up sharing certain personal communications with a different service provider?

All of this also needs to be compatible with existing message franking protocols~\cite{grubbs2017message} (and their metadata-private counterpart, asymmetric message franking~\cite{tyagi2019asymmetric}). Message franking prevents users from faking abuse reports by cryptographically stamping each message; it has been deployed in Meta messaging services for several years now~\cite{messenger_whitepaper}.

\vspace{.7em}
\noindent\textbf{Metadata:} Since user reporting is inherently retroactive (i.e., the harmful content has already been sent), service providers also rely extensively on metadata to monitor unusual communication patterns in real time. This is most obviously relevant for spam, where a service provider can detect unusual volumes or destinations of messages, but profile or chat descriptions can also be very useful for fighting abuse. WhatsApp bans over 300,000 accounts each month for CSAM sharing based on this approach~\cite{whatsapp_cei}.

From a privacy standpoint, this dependence on metadata raises numerous questions around what data would be viewable by each service provider in an interoperable communication. While the DMA specifies that only personal data that is ``strictly necessary'' for ``effective interoperability'' can be shared between providers, this can be quite expansive given providers' reliance on metadata. On the other hand, if a provider is overly limited in what data they can collect, this could have adverse effects on efforts to fight spam and CSAM, negatively impacting the user experience. Len et al.~\cite{len2023interoperability} propose to resolve this tension by dividing the responsibility for spam filtering between the sender's provider and the recipient's provider, with the expectation that the sender's provider will do some degree of metadata-based filtering prior to passing the message with metadata removed along to the recipient (or opt not to send the message at all if they deem it to be spam). This would require each platform to rely in large part on an external third-party (the sender provider) for spam filtering, and represents a fundamental reimagining of which parties in a communication channel are responsible for moderating content by shifting much of the filtering to the sender. The feasibility of this proposal in practice will depend on as-yet-undetermined European Commission standards for when and how quickly one provider is able to break off interoperability with another if the level of filtering proves inadequate.

And it cannot be assumed that only two platforms would be involved. If Alice and Bob are on Signal, and Bob adds Carol and Dave on WhatsApp, and Dave then adds Eolina and Firoz on Telegram, and so on, do we maintain message forwarding limits, source tracing, and other moderation techniques end-to-end? WhatsApp, for instance, imposes restrictions on how many times a message can be forwarded to limit viral spread (as this is abused to promote disinformation, such as election-related hoaxes)~\cite{whatsapp2023forwarding}. Without an inter-provider limit, spam and disinformation can be laundered in just the same way that drug gangs launder their proceeds by sending them along a chain of bitcoin exchanges. But cross-platform forwarding limits will only work if everyone keeps proper records in compatible ways. If you believe that's going to happen, we have a bridge we'd like to sell you.

\subsubsection{Content Detection Schemes}
\label{section:content_detection}
\hfill\\
\vspace{-.5em}

\noindent While user reporting and metadata analysis are already deployed by most platforms, some may additionally deploy other content moderation schemes which arguably undermine the confidentiality and end-to-end encrypted properties of the messages, whether of their own volition or because they are under a government mandate to do so.

\vspace{.7em}
\noindent\textbf{Content Scanning:}
Suppose a platform using some kind of automated content detection system (such as perceptual hashing~\cite{kulshrestha2021identifying}, which can detect known harmful content in encrypted communications but has been shown to be vulnerable to attack~\cite{jain2022adversarial,prokos2023squint}) requests to interoperate with a service that has consciously opted not to use such a system. Will the latter be allowed to refuse this request on the grounds that it would compromise the level of security they provide to users?

And perhaps it is the other way around, where it is the gatekeeper that has deployed some type of client-side scanning mechanism, either voluntarily or under a government mandate. Interoperability can be co-opted by governments to undermine end-to-end encryption in various ways. There is a clear risk, given the Child Sex Abuse (``Chat Control") Regulation before the European Parliament~\cite{eu2022csam} and the Online Safety Bill before the British one, that the agencies will mandate client-side scanning on precisely those large platforms on which the interoperability mandate falls. In that case, the mandated client-side bridge becomes a scanning gateway that is in effect under government control. Recent history is littered with examples where law enforcement and the technical community held rather different notions of what violates an existing level of security \cite{abelson1997risks,abelson2015keys,levy2018principles,abelson2021bugs,levy2022thoughts,muffett2022glossary}.

\vspace{.7em}
\noindent\textbf{Traceability:}
Message traceability is the idea that platforms should be able to identify the source of a heavily forwarded message~\cite{tyagi2019traceback}. WhatsApp and numerous civil society organizations have spoken out against traceability, arguing that it breaks end-to-end encryption by identifying content users have shared without the consent of the message sender or recipient(s)~\cite{whatsapp2023traceability}. Will WhatsApp be required to interoperate with a platform that has deployed some form of message traceability?

Broadly, will the European Commission consider systems using content scanning, traceability, or other moderation techniques to endanger the integrity of a service that does not deploy these techniques, given that they are themselves pursuing such mandates? More to the point, how would a gatekeeper even know whether the requesting service has deployed these or other schemes? Meredith Whittaker, President of the Signal Foundation, has alluded to this challenge, stating that while Signal is open to the general notion of interoperability, they would need to ensure there are no ``tricks on the backend" to compromise security or privacy~\cite{angwin2023whittaker}.

\subsection{User Interface Design}
\label{section:user_interface}
Interface design is critical if messaging interoperability is to enhance, rather than degrade, the user experience. We have already seen that interoperable communication inherently presents a reduction in security and privacy compared to communication taking place on a single service. A panelist at a recent DMA stakeholder workshop, however, suggested that ``we have failed if we have to inform the user of something changing''~\cite{ec2023stakeholders}. This is a well-intentioned but misguided notion that understates the data-sharing implications of interoperability and the depth of the privacy protections promised by platforms. Privacy-preserving cryptographic techniques can only take us so far. Critically, some data \textit{cannot} be hidden in order to allow for sufficiently accurate spam moderation. We will need clear and unambiguous ways of informing the user that their data and messages are leaving their service, and, by extension, that the security and privacy guarantees and features of their platform may no longer apply.

\subsubsection{App Selection Interface}
\hfill\\
\vspace{-.5em}

\noindent The Matrix Foundation, Cory Doctorow, and others have drafted preliminary mockups of what the interface for selecting the message recipient's service might look like \cite{hodgson2023blog,doctorow2022facebook}, with Matrix likening it to choosing which application to use to open a file on an OS (``Open with...''). If a user opts to be discoverable on a large number of services (say, $n > 5$), we risk the interface becoming cluttered, but we consider this a comparatively minor problem since a limited number of services will be designated gatekeepers by the EU to begin with. Rescorla~\cite{rescorla2022blog} suggested the example of email provider selection, where a user is shown a window containing the five or so most common email providers along with a freeform ``other'' option to enter an alternative provider. Such an interface was mandated by the EU for default browser selection on Windows PCs from 2009--2014, with a random choice order, after an antitrust finding against Microsoft's Internet Explorer.

There are still many open design questions around user discovery that determine what would be displayed to the user. If Bob is discoverable on only one service, should Alice's client default to sending the message to Bob on that service? Or should Alice be shown a pop-up window with just one option that she still needs to manually select in order to ensure she understands and consents to starting a communication channel with Bob's service provider? If Bob is discoverable on multiple services, and one of them happens to be the same service that Alice uses, should the chat default to Alice's service? Suppose Bob has two distinct ongoing communication channels with Alice through different apps Bob uses. Will the messages be intermingled within the same interface (as is the case with iMessage's SMS interoperability), or will they be shown to the user as two separate chats?

If Bob designates a particular platform as his preferred service, would Alice's interface indicate Bob's preferred service (for instance, via a visual nudge to choose it), or would the chat simply default to it? Would Bob be able to change this after the fact, or would they have to start a new conversation? What if Bob wants his default service to be whatever service Alice is using, and to choose a different one only if Bob is not on Alice's service? Most importantly, can individual platforms make different decisions in response to these questions, or do they need to establish a set of agreed standards for the interoperable user experience? Design questions like these will have an enormous impact on the user experience.

\subsubsection{Communicating Changes in Security Guarantees}
\hfill\\
\vspace{-.5em}

\noindent
Effectively communicating security and privacy risks to the user is arguably a more difficult problem than designing a new cryptographic protocol for data exchange. We can draw on decades of security usability research showing time and again that users struggle to comprehend and act on data access requests \cite{felt2012android,king2011privacy,santos2021cookie} and security warnings \cite{sunshine2009crying,schroder2016signal,reeder2018experience,petelka2019put}. Among other takeaways, researchers have concluded that to be effective, the text of warnings need to communicate clear and specific risks~\cite{stransky2021limited}. The user would presumably be given a textual warning with an option to read further details at the moment they opt to be discoverable by an alternative service.

The open question is what distinction, if any, would be shown on a chat-by-chat basis. Matrix has compared this to the WhatsApp Business API \cite{cardozo2020business,ec2023stakeholders}, which is not end-to-end encrypted when a business uses a third-party vendor (Meta included) to host their API. But while ordinary WhatsApp chats have a small bubble at the top of their chat window informing them that their communications are encrypted end-to-end, a chat passing through the business API simply makes no mention of encryption at all. In other words, WhatsApp ``informs'' users of the reduction in security through the absence of a typical security indicator, a design which is likely ineffective in building user comprehension. To avoid inadvertent downgrade attacks, it will have to be abundantly clear to a user whether a given conversation is taking place within their own service only, or across multiple third-party services \cite{abu2018exploring}. And visual indicators are no silver bullet: Signal recently removed SMS support for Android, citing, among other reasons, the need to avoid inadvertent user confusion given that SMS and Signal messages were both sent in the same interface. As Signal put it, they “can only do so much on the design side to prevent such misunderstandings” \cite{nina2022sms}.

We know from existing usability research that whether a user takes a warning seriously depends heavily on the design \cite{utz2019informed}. In one example back in 2013, Akhawe et al.~\cite{akhawe2013alice} found that Mozilla Firefox's SSL warnings were significantly more effective than those in other browsers due to variations in design: around 70\% of users clicked through Chrome’s SSL warnings, while only one-third clicked through Firefox’s warnings. Chrome has since invested heavily in redesigning their SSL warnings based on what Felt et al.~\cite{felt2015improving} termed ``opinionated design'', meaning that the user should be nudged towards the option perceived by the designer to be superior through visual cues (instead of text explanations). In practice, this is typically accomplished by varying colors to encourage the user to choose the safer option or requiring the user to click through an extra window before advancing. Of course, in industry such ideas have been co-opted to create interfaces that nudge users to select an option that is in the company's interest---the more well-known term for these design strategies is now ``dark patterns''~\cite{gray2018dark}.

\subsubsection{Blue Bubbles and Green Bubbles}
\hfill\\
\vspace{-.5em}

\noindent
Fortunately, we already have a real-world case study demonstrating the impact of interface design choices on user decisions in an interoperable setting. While Apple's iMessage interoperates with SMS/MMS, Apple uses visual contrasts to make the difference abundantly clear to the user. When two iPhone users communicate through Apple’s default Messages app, the sender’s messages appear blue (indicating that the communication is taking place over Apple’s iMessage). In contrast, while an iPhone user is able to use the same interface to talk to an Android user, the iPhone user’s sent messages now appear green, indicating that the messages are being sent over SMS. This distinction has given rise to social pressure to use an iPhone over an Android phone, thereby further consolidating Apple's market control, particularly over younger generations~\cite{mcgee2023apple,higgins2022apple}. In the words of one user, ``If that bubble pops up green, I'm not replying"~\cite{mcgee2023apple}. 

While the bubble color is not the only difference the user experiences (green bubbles lack certain iMessage features like emoji reactions, group naming, typing indicators, etc.), the rapid propagation of the blue versus green bubbles phenomenon throughout popular culture demonstrates just how effective these types of visual cues can be in branding one option as ``good'' and the other ``bad''---indeed, from Android's perspective, perhaps a little too effective~\cite{cole2022bubbles}.

The ongoing bubble war dispels any notion that interoperability on its own will defeat network effects. Far from opening up Apple's ``walled garden'', iMessage/SMS interoperability appears to have further solidified Apple's market power. At the same time, however, Apple has an obligation to its users to inform them of the difference for fundamental security reasons. The same is true even when two E2EE services interoperate: as Meredith Whittaker recently pointed out, simply being end-to-end encrypted is not ``an end in itself''~\cite{whittaker2023tweet}. Messaging services have widely varying privacy policies, cloud backup schemes, jurisdictions in which they operate, receptiveness to client-side scanning, levels of cooperation with foreign governments, and so on. Rather, the goal must be to preserve a broader understanding of privacy and situational awareness of possible attacks. The wicked question here, and one for which we have no answer, is how to convey accurately and clearly to the user what is happening when they opt to interoperate with another service, while not needlessly discouraging them from doing so.


\section{Discussion}
Having explored in depth what an interoperable messaging solution might look like, we can see that the core message communication protocol is just the beginning. Interoperable systems will also need protocols to support the many other features that make secure communication possible, including cross-platform user authentication and tracking data exchanges between platforms. A harbinger of the technical difficulty may be the fact that Meta has been trying to interoperate WhatsApp and Facebook Messenger since 2019 -- and this is for two services owned and operated by the same company, which has a strong business incentive to make it work~\cite{zuckerberg2019privacy,azhar2021regulate}. Meta's President of Global Affairs, Nick Clegg, acknowledged in a 2021 interview that inter-platform interoperability is ``taking us a lot longer than we initially thought''~\cite{azhar2021regulate}.

In any serious discussion of security and privacy trade-offs, the yardstick needs to be how much the attack surface has increased compared to existing systems. For instance, disappearing messages, a feature allowing a user to set messages to be automatically deleted after some specified period of time, have been held up as an example of a feature that is difficult to ensure in an interoperable service~\cite{newton2022cathcart,brown2022whitepaper}. Even if both services ostensibly offer the feature, each has no guarantee that the other has actually complied with the deletion request. When the subject came up at the European Commission's stakeholder workshop, the panelists' response was that a user never really knows what happens to their communications once they've left their device anyway---the message recipient may take a screenshot, have malware on their device, and so on~\cite{ec2023stakeholders}. This rejoinder dodges the reality that interoperability would add a substantial new attack vector in the form of the interoperating service. That disappearing messages, and indeed many of the challenges discussed throughout the paper, are already concerns in existing systems is no reason to make things worse.

Finally, interoperability will have to respect users' moral or personal-security choices. Some users so dislike Meta that they refuse to use WhatsApp, and demand that their friends use Signal instead. Other users, for example in Ukraine, consider Telegram to be compromised by Russian state agencies and refuse to touch it~\cite{loucaides2023telegram}. It will have to be obvious to everyone in a group when someone joining the group is acting as a gateway to another service, and the notification will have to be transitive, so that every group member can see whether their communications will end up in a system they do not trust. But here again, there are trade-offs between security and usability: if all users in a group have to approve the addition of a new member who comes from an odd network, the effect will be like forcing people to opt in to ad tracking -- it won't happen at scale. 


Interoperability without robust moderation and interface design to make platforms pleasant to use is a nonstarter. Giving users a choice between platforms without giving them a platform they would want to spend time on is no choice at all. The challenges in this space, though, are all the more reason for researchers to devote effort and attention to it, to ensure that users continue to have secure channels of communication in an interoperable world.

\section{Acknowledgements}
We thank Alastair Beresford, Ian Brown, Jon Crowcroft, Phillip Hallam-Baker, Daniel Hugenroth, Alec Muffett, Sam Smith, and Michael Specter for valuable feedback and/or discussions. These contributors do not necessarily agree with the arguments presented here.

%
%
\bibliographystyle{splncs04}
\typeout{}
\bibliography{ref}
\end{document}